\documentclass[a4paper]{jpconf}
\usepackage{graphicx}
\begin{document}
\title{Evolution of the decay mechanisms in central collisions of $Xe$ + $Sn$ from $E/A$ = 8 to 29 $MeV$ }

\author {A. Chbihi$^1 $, L. Manduci$^2 $, J.~Moisan$^1 $, E.~Bonnet$^1 $, J.~D.~Frankland$^1 $, R.~Roy$^3 $, G.~Verde$^4 $ }

\address {$^1 $ GANIL, CEA et IN2P3-CNRS, B.P.~5027, F-14076 Caen Cedex, France.}
\address {$^2 $ Ecole des Applications Militaires de l'Energie Atomique, BP 19, 50115, Cherbourg, France.}
\address{$^3 $  Laboratoire de Physique Nucl\'eaire, Universit\'e Laval, Qu\'ebec, Canada G1K 7P4.}
\address{$^4 $ Istituto Nazionale di Fisica Nucleare, Sezione di Catania, 64 Via Santa Sofia, I-95123, Catania, Italy}

\ead{chbihi@ganil.fr}

\begin{abstract}
Collisions of Xe+Sn at beam energies of $E/A$ = 8 to 29 $MeV$ and leading to fusion-like heavy residues are studied using the $4\pi$ INDRA multidetector. The fusion cross section was measured and shows a maximum at $E/A$ = 18-20 $MeV$. A decomposition into four exit-channels consisting of the number of heavy fragments produced in central collisions has been made. Their relative yields are measured as a function of the incident beam energy.  The energy spectra of light charged particles (LCP) in coincidence with the fragments of each exit-channel have been analyzed. They reveal that a composite system is formed, it is highly excited and first decays by emitting light particles and then may breakup into 2- or many- fragments or survives as an evaporative residue. A quantitative estimation of this primary emission is given and compared to the secondary decay of the fragments. These analyses indicate that most of the evaporative LCP precede not only fission but also breakup into several fragments.            
\end{abstract}

\section{Introduction}

Reactions between heavy nuclei at low energy above the barrier are dominated by binary inelastic collisions \cite{sch78,gla83,ste95,chimera}. Even the most central collisions ($low-l$ partial waves) lead to highly-damped two-body exit channels, and no more fusion is observed. The attractive pocket in the internuclear potential disappears, as, for large enough values of the charge product of projectile and target,($Z_p$$Z_t$), the bombarding energy necessary to overcome the Coulomb barrier is such that the repulsive  potential is always stronger than the attractive nuclear one. According to the prediction of the classical potential model of Bass, applied to the fusion of heavy nuclei,  the limiting value for fusion is given roughly by the product $Z_p$$Z_t$ = 2700-2800 \cite{bas74,tam75}. 

However, the study of central collisions of heavy nuclei in the intermediate energy regime ($E/A$ = 20-100 $MeV$) has shown the existence of compact nuclear sources for which multifragmentation is increasingly the dominant decay channel. As an example of compact sources claimed for a few heavy systems around the Fermi energy : $Xe$ + $Sn$ at $E/A$ = 32-50 $MeV$, $Gd$ + $U$ at $E/A$ = 36 $MeV$, $Au$ + $Au$ at $E/A$ = 35 $MeV$ and $Pb$ + $Au$ at $E/A$ = 29 $MeV$\cite{mar97, hud03,fra011,fra012,neb99,ago96}.    For these systems, central collisions have been kinematically consistent with the breakup of large composite systems. The cross section of these events has been estimated around 10 to 100 $mb$. Moreover, the study of the system  $Xe$ + $Sn$ at $E/A$ = 25 $MeV$ has revealed the survival of  heavy residues with $Z$ $\approx$ 70 suggesting the formation of excited compound nuclei which decay by emission of neutrons, light charged particles (LCP) and a few intermediate fragments (IMF)\cite{fra02}. 

The aim of this work is to study the evolution of the reaction mechanism occurring in heavy ion collisions from low energies dominated by binary exit-channels to intermediate energies dominated by multi-fragment emission from a compact-source. 
The transition between the two extreme pictures should clarify the mechanism responsible for the formation of such a source. Is it fusion or massive transfer between the two heavy fragments ? Does the attractive nuclear potential overcome the repulsive one ? If so, at which energy ? What is the fraction of fused nuclei that decay by fission, 3- or more-fragment break-up? What is the size of the heaviest fragment which can survive fission ?  

In this contribution, we will provide an answer to some of these questions, by reporting on measurements  performed with the INDRA multidetector of quasi-symmetric heavy system $^{129}Xe$+$^{nat}Sn$ having  $Z_P$$Z_T$ at the limit (2700)  at the bombarding energy $E/A$ = 8, 12, 15, 18, 20, 25, 27 and 29 $MeV$; and $^{129}Xe$+$^{197}Au$ with $Z_P$$Z_T$ = 4266 at $E/A$ = 15 $MeV$. We will present in section 2 the experimental methods and the event selection. The fusion cross section and the relative yield of the different exit-channels will be given in section 3.  The size of the primary source and the secondary decay are presented in section 4, and then we will conclude.

\section{Experimental procedures and event selection}

The experiment has been performed at GANIL facility. It consists of the coupling of two main cyclotrons CSS1 and CSS2. This combination does not allow to obtain the incident energy range of E/A = 8  to 20 MeV for the $Xe$ beam. Therefore the beam of $^{129}Xe$ was first accelerated to E/A = 27 MeV and then slowed down, using carbon-degrader foil with different thicknesses, to $E/A$ = 25, 20, 18, 15, 12 and 8 MeV. The charge state of the primary beam was 40+, after the degrader the Xe beam, as expected,  had a wide distribution of charge states. We therefore used the alpha-spectrometer of GANIL in order to select only one charge state. The  B$\rho$ setting of the spectrometer was optimized for each incident energy.  However for the two lowest energies, $E/A$ = 8 and 12 $MeV$, more than one charge state were transmitted given uncertainties on these two incident energies. This uncertainty is estimated to be $\delta$$E/A$ = 0.5 (0.2) MeV for the $E/A$ = 8 (12) MeV beam respectively.   The energy $E/A$ = 29 MeV was obtained by direct tuning.   
  
The beam of $^{129}$Xe was used to bombard a self-supporting 350 $\mu$g/cm$^2$-thick $^{nat}$Sn (or 100$\mu$g/cm$^2$-thick $^{197}Au$) target
placed inside the INDRA detector array\cite{pou95,pou96}. This charged products detector covers 90\% of the $4\pi$ solid angle. The total number of detection cells is 336 arranged according to 17 rings centered on the beam axis. The first ring (2$^o$ - 3$^o$) is made of 12 telescopes composed of 300 $\mu$m silicon wafer and CsI(Tl) scintillator (14 cm thick). Rings 2 to 9 (3$^o$ - 45$^o$) have 12 or 24 three-member detection telescopes : a gas-ionization chamber, a 300 or 150 $\mu$m silicon wafer and CsI(Tl) scintillator (14 to 10 cm thick) coupled to a photomultiplier tube. Rings 10 to 17 (45$^o$ - 176$^o$) are composed of 24, 16 or 8 two-member telescopes : a gas-ionization chamber and a CsI(Tl) scintillator of 8, 6 or 5 cm thickness. Fragment identification thresholds are around 0.5 $MeV$/nucleon for the lightest (Z $\approx$ 10) and 1.5 $MeV$/nucleon for the heaviest (Z $\approx$ 50). An extrapolation  using the energy loss tables makes possible an identification up to charge $Z$ = 80 with a resolution of 5 charge units.  

Events were recorded with an on-line trigger requiring 2 or more independent
detectors to be hit in coincidence. In the off-line analysis a software trigger
requiring 2 or more charged products in each event was applied, giving a data
sample of \mbox{$5-10\times 10^6$} events.

\begin{figure}[h]
\includegraphics[width=36pc]{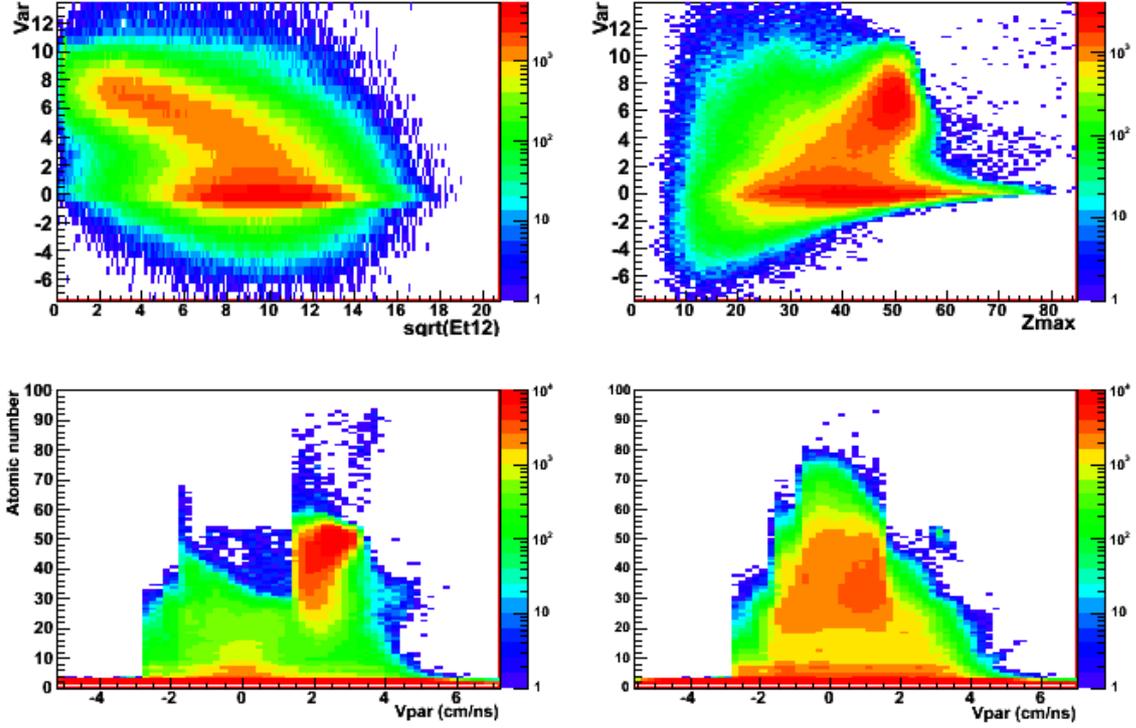}
\caption{\label{epseudo} For the $^{129}Xe$+$^{nat}Sn$ at $E/A$ = 15 $MeV$ : $E_{pseudo}$ as a function of $\sqrt{E_{\perp}^{12}}$ (top left panel) and as function of the heaviest fragment in the event $Z_{max}$ (top right panel). Atomic number of the fragments (bottom panels) vs their $V_{\parallel}^{cm}$ for events selected outside window -2 $\leq$ $E_{pseudo}$$\leq$ 2 (left) and inside the window(right).}
\end{figure}

To select the central collision, we have used a kinematic global variable applied to the heaviest fragment in the event and defined as : 
\begin{equation}
E_{pseudo} = V_{\parallel}^{2} -  \frac{1}{2} V_{\perp}^{2}
\end{equation}
where $V_{\parallel}$ and $V_{\perp}$ are the parallel and transverse velocity of the heaviest fragment of the event in the center of mass frame. This variable\cite{bou}, $E_{pseudo}$ that we will call pseudo-energy, amplifies the separation between the projectile-like fragment component and residue produced at rest in the center of mass frame. If the heaviest fragment is emitted exactly at the center of mass velocity, the value of $E_{pseudo}$ is zero, if it is a projectile-like fragment its value will be much greater than zero.  $E_{pseudo}$ is plotted in  figure \ref{epseudo} (upper panel-left) as a function of the square root of the transverse energy of light charged particles (LCP), $\sqrt{E_{\perp}^{12}}$, for the reaction $^{129}Xe$+$^{nat}Sn$ at $E/A$ = 15 $MeV$ .  This latter variable gives the degree of centrality. Two components  are clearly shown one centered at 
$E_{pseudo}$ = 0 (y-axis) and at significantly high $\sqrt{E_{\perp}^{12}}$ ($\approx$ 10), whereas  the ridge of the second  component decreases from  $E_{pseudo}$ $\approx$  7 at low $\sqrt{E_{\perp}^{12}}$ $\approx$ 3 toward $E_{pseudo}$ $\approx$  2-3 at high $\sqrt{E_{\perp}^{12}}$ $\approx$ 10. This evolution reflects the increasing energy dissipation of the projectile and target from peripheral to central collisions. A deep valley  of the distribution can be observed at $E_{pseudo}$ $\approx$  2 which facilitates the separation between binary events (deep-inelastic collisions) and  central collisions (a good candidate for fusion). The same behavior can be observed in upper-right panel where  $E_{pseudo}$ is plotted vs the atomic number of the heaviest fragment ($Z_{max}$). Beside the projectile-like component centered at $Z_{max}$ = 54, one can observe  in the window -2 $\leq$ $E_{pseudo}$$\leq$ 2 a broad distribution of $Z_{max}$ centered at 40 and ranging from 20 to 80. The effect of this global variable on all reaction products is shown in figure \ref{epseudo} (lower panels) where are presented the atomic number  of the fragments as a function of their cm-parallel velocity ($V_{\parallel}^{cm}$) with the condition outside the window -2 $\leq$ $E_{pseudo}$$\leq$ 2 (left panel ) and inside the window (right panel). Projectile and target like components are clearly seen when selecting outside the window while inside the window, products with charge up to 80 are centered at the center of mass velocity when events with -2 $\leq$ $E_{pseudo}$$\leq$ 2 are selected. 

The advantage of this selection is that no completeness of the events is needed. Therefore the cross section of central collisions can be deduced. However we should correct for the detection efficiency of the heaviest fragment.

\section{Fusion cross section and relative yield of exit channels}

\begin{figure}[h]
\begin{minipage}{20pc}
\includegraphics[width=20pc]{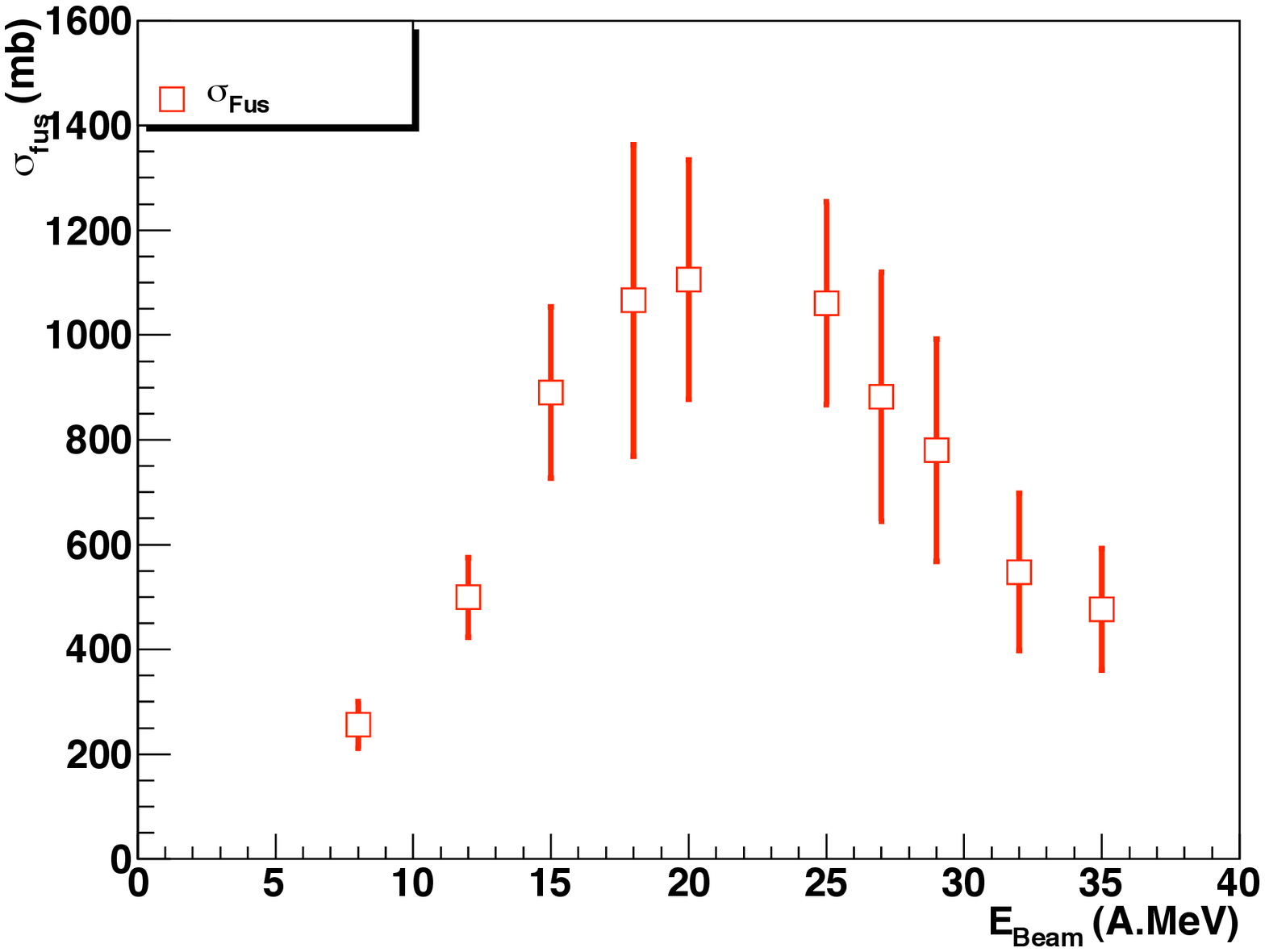}
\caption{\label{cxsec}Cross section of central collisions not corrected for efficiency.}
\end{minipage}\hspace{2pc}%
\begin{minipage}{20pc}
\includegraphics[width=20pc]{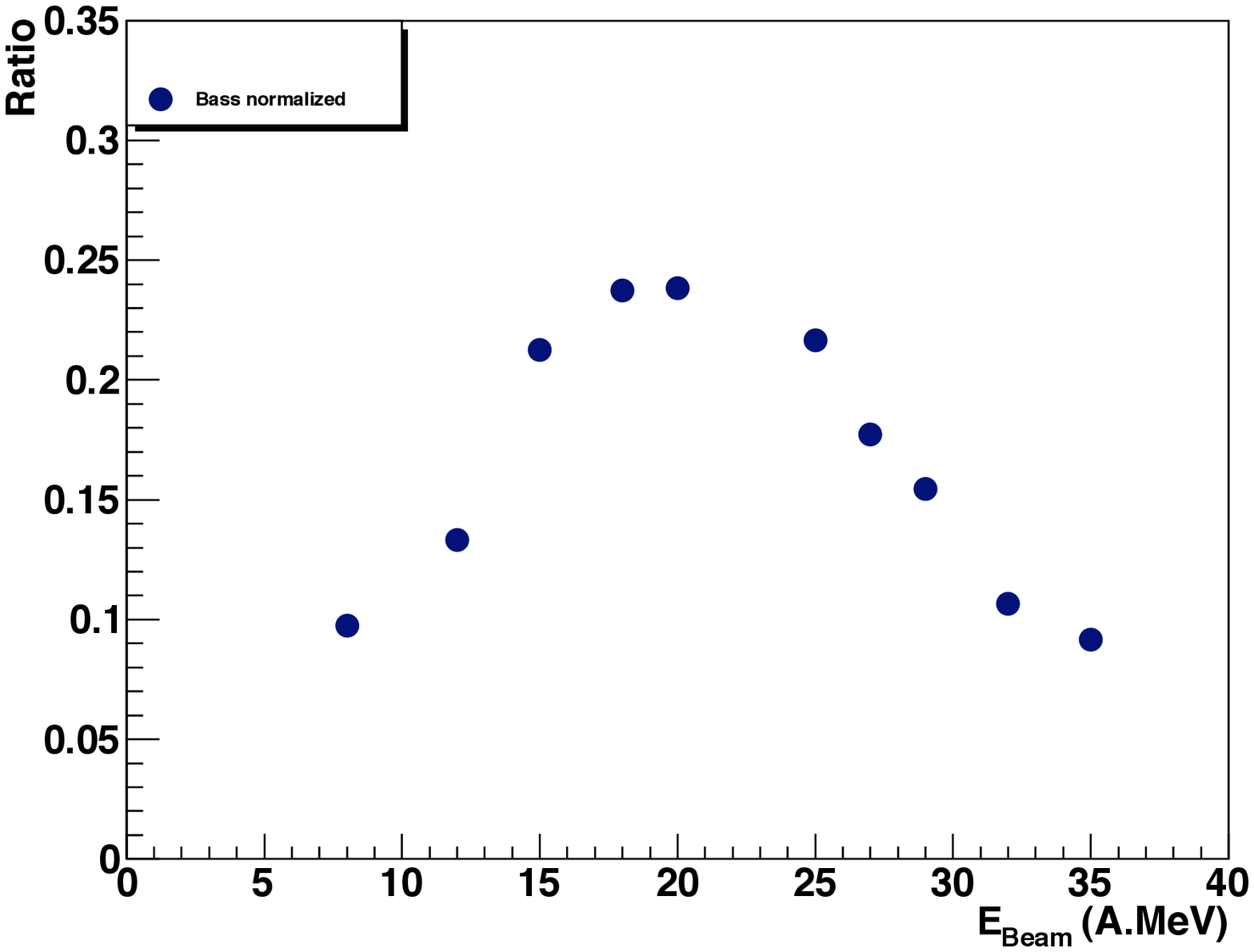}
\caption{\label{cxsecn}Cross section of central collision normalized to the reaction cross section from Bass prediction\cite{bas74}.}
\end{minipage} 
\end{figure}

We used the $E_{pseudo}$ variable to  select the central collisions with a minimum bias. The windows of pseudo-energy applied to each bombarding energy are tabulated in table \ref{tab1}. 
Within this selection the cross section of central collisions is shown in figure \ref{cxsec} as measured and in \ref{cxsecn} normalized to the total reaction cross section estimated from Bass calculation. No correction for efficiency has been applied. Both distributions show a bell-shape, the maximum is located at $E/A$ $\approx$ 18-20 $MeV$. At this maximum the cross section value that we obtain is $\approx$ 1.1 $mb$  corresponding to 25\% of the total reaction cross section estimated from Bass predictions \cite{bas74}. If one assumes that the cross section of central collision is fusion cross section, the behavior  of the normalized cross section (figure \ref{cxsecn}) does not follow the trend of the fusion cross sections of asymmetric systems found in the literature \cite{eud12}. For the asymmetric systems the normalized fusion cross section decreases with the incident energy.  We have to notice also that the incident energy corresponding to the maximum normalized cross section  ($E/A$ = 18-20 MeV) is more than 4 times the Coulomb barrier. 
\begin{center}
\begin{table}[h]
\caption{\label{tab1}Limits  of pseudo-energy applied to each bombarding energy for the  $^{129}Xe$+$^{nat}Sn$ systems and the cross section of central collisions}
\centering
\begin{tabular}{@{}*{7}{l}}
\br
$E_i$/$A$ ($MeV$)&$E_{pseudo}^{min}$&$E_{pseudo}^{max}$&$\sigma$ (mb) &$\delta$$\sigma$ (mb) \\
\mr
\verb"8"&-1.4  &0.4 &256 &45\\
\verb"12"&-1.0 & 0.8 &500 &76\\
\verb"15"&-1.0 &1.0 &891 &164\\
\verb"18"&-1.5 &1.5 &1066 &297\\
\verb"20"&-1.5 &2.0 &1106 &228\\
\verb"25"&-2.0 &2.0 &1061 &193\\
\verb"27"&-1.0 &2.0 &883 &237\\
\verb"29"&-1.0 &2.0 &780 &213\\
\verb"32"&-1.0 &2.0 &548 &150\\
\verb"35"&-1.0 &2.0 &477 &116\\
\br
\end{tabular}
\end{table}
\end{center}

\begin{figure}[h]
\includegraphics[width=40pc]{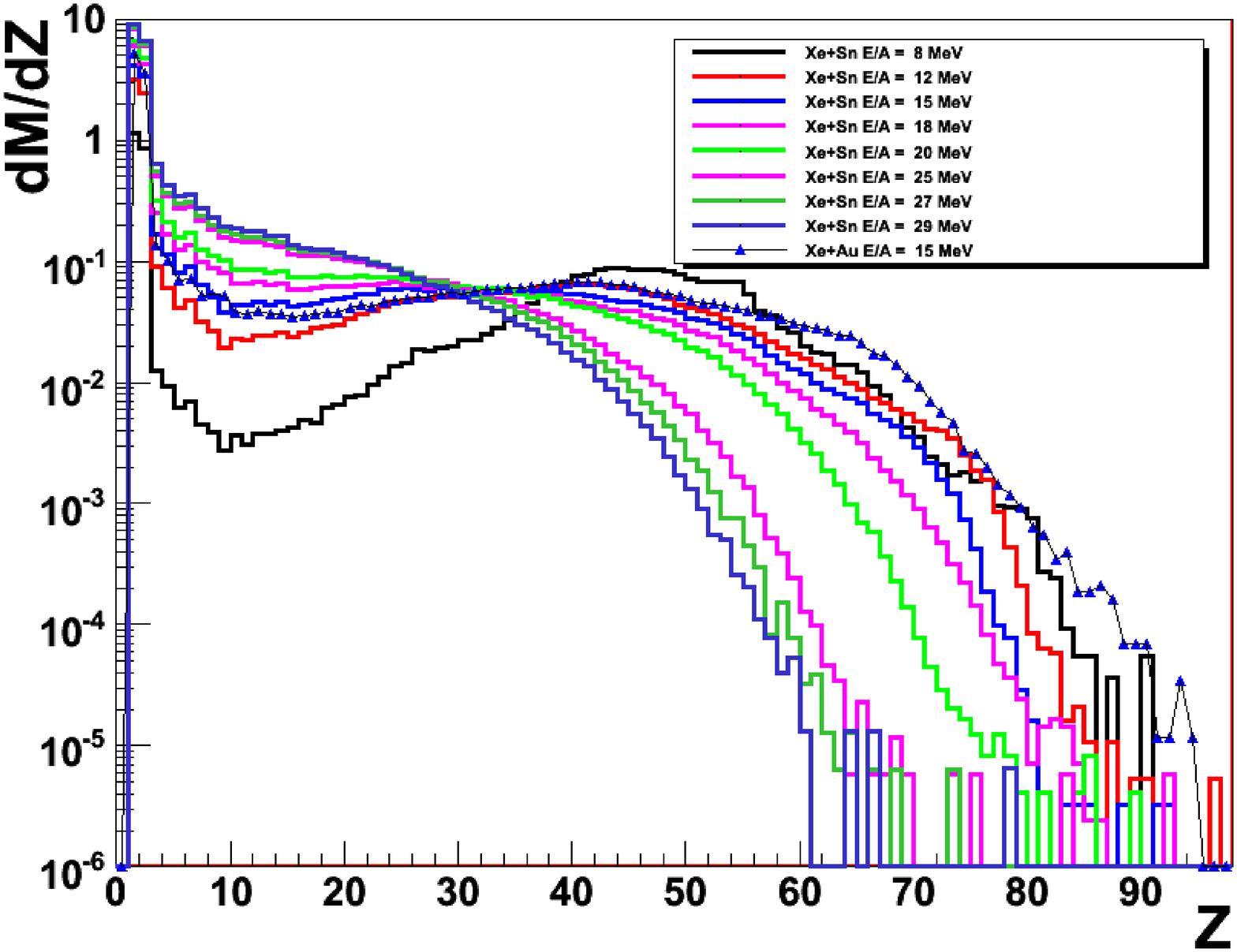}
\caption{\label{zdis} Charge distributions produced in central collisions of $^{129}Xe$+$^{nat}Sn$ at $E/A$ = 8-29 $MeV$ and  $^{129}Xe$+$^{197}Au$ at $E/A$ = 15 $MeV$.}
\end{figure}

The charge distributions produced in central collisions of  the $^{129}Xe$+$^{nat}Sn$ at $E/A$ = 8-29 $MeV$ and  $^{129}Xe$+$^{197}Au$ at $E/A$ = 15 $MeV$ are presented in figure \ref{zdis}. They are normalized to the number of events of each incident energy. therefore, the y-axis represents the multiplicity of each charge. The distributions are very broad, they cover the atomic number range from $Z$ = 1 to $Z$ $\approx$ 90. The tail of the charge distribution extends toward higher and higher values when the incident energy decreases.  Heavy residues with  $Z  \geq$ 80, which represent 77\% of the total charge of the Xe+Sn and   60\% for the Xe+Au, survive with significant cross section. One observes also a high multiplicity of LCP which decrease rapidly with increasing charge and then presents a deep minimum in the distribution at $Z$ = 10. This minima shifts towards a lower Z with increasing incident energy and vanishes at the highest energy. In the following, we will call the value of this minimum $Z_{min}$. Its value is defined as~: $Z_{min}$=10 for the energy range $E/A$ = 8 to 15 MeV and $Z_{min}$ = 5 for the higher energies.  In low energy regime, this shape of the charge distribution is an intrinsic characteristic of heavy residue formation by emission of LCP and small clusters. The residues can be compound nuclei or fission fragments. This feature is very well reproduced by statistical models. We therefore define two regions in the charge distribution~: one with  $Z  \geq$ $Z_{min}$ corresponding to residues; the second, with  $Z  <$ $Z_{min}$  corresponding to LCP's and light clusters which result from evaporation.  

\begin{figure}[h]
\includegraphics[width=36pc]{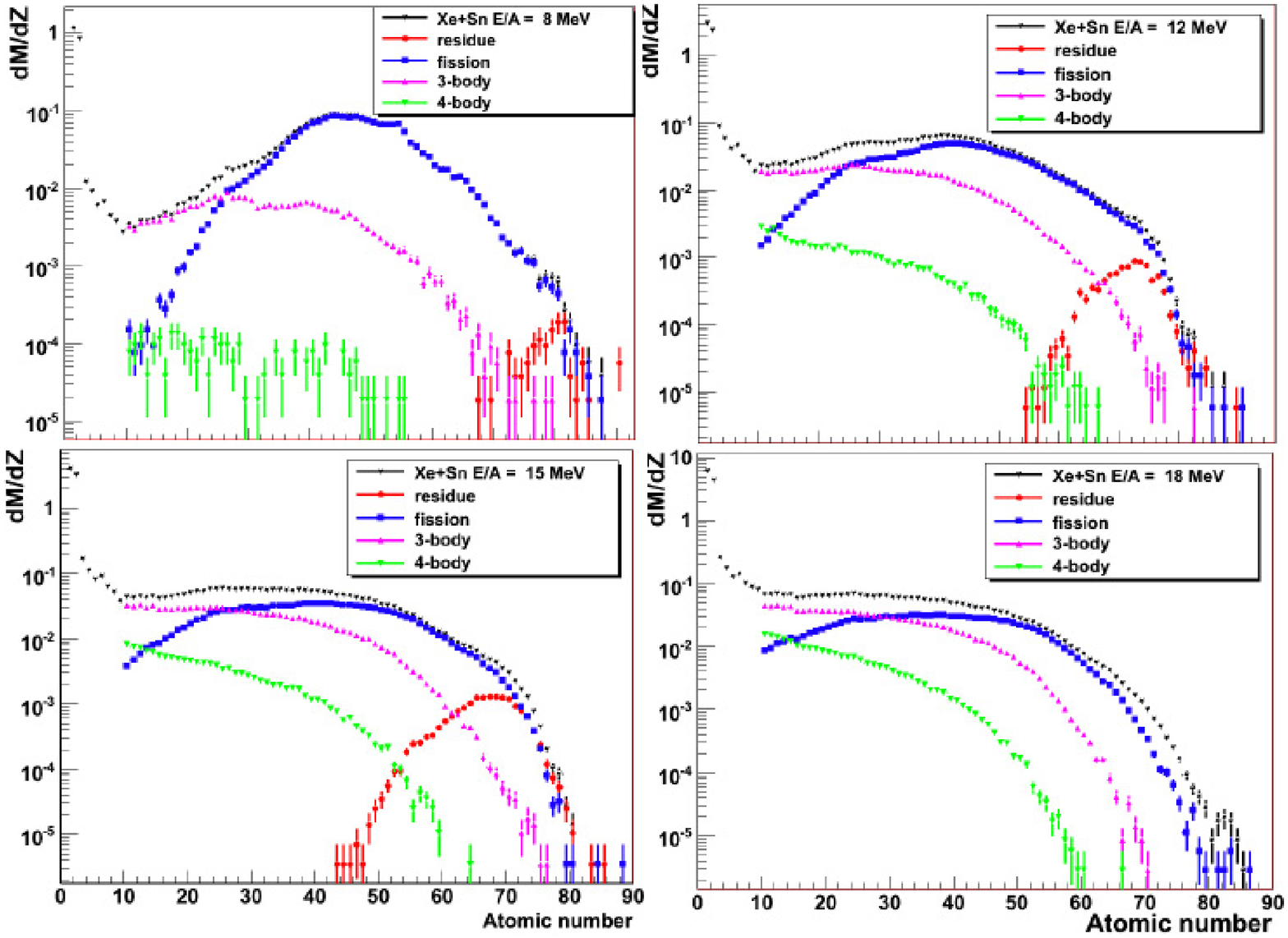}
\caption{\label{exitch}Exit channel decomposition for the central collisions of $^{129}Xe$+$^{nat}Sn$ at $E/A$ = 8, 12, 15 and 18 $MeV$ systems. No efficiency correction has been applied.}
\end{figure}

\begin{figure}[h]
\includegraphics[width=40pc]{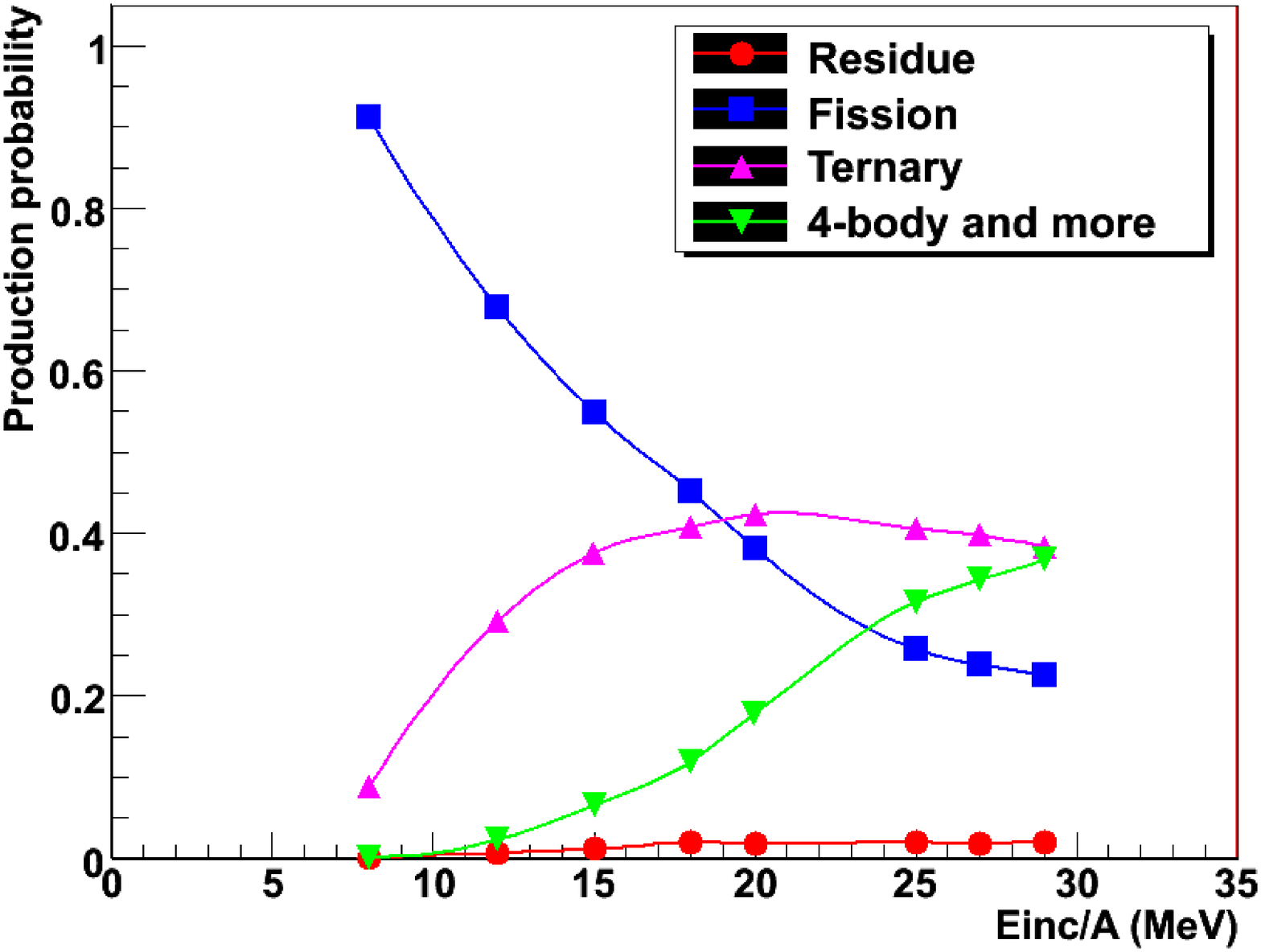}
\caption{\label{decomp}Production rate of the four exit channels vs the incident energy for the central collisions of $^{129}Xe$+$^{nat}Sn$. No efficiency correction has been applied}
\end{figure}

We have identified 4 channels differing by the number of fragments with charge $Z\geq$ $Z_{min}$: 1, 2, 3 and 4 fragments. To take into account the efficiency of the detector, at least partially, only events having total charge $Z_{tot} \geq$ 83 have been kept. Figure \ref{exitch}  shows the repartition of the charge distribution among the 4-exit-channels for the systems $^{129}Xe$+$^{nat}Sn$ at $E/A$ = 8, 12, 15 and 18 $MeV$. One can observe the evolution of the exit-channels as a function of the incident energy. The distributions are broad, in particular for the 2-fragment exit channel that we call "fission". They extend from their lower limit $Z_{min}$  up to $Z = $80-82. The  one-fragment exit-channel is produced with very low cross section at the lowest energy, it increases at $E/A$ = 12 and 15 $MeV$ and disappears at 18 $MeV$. The relative probabilities of the four exit channels as a function of the incident energy are shown in figure \ref{decomp}. The fission represents more than 90\% at the lowest energy. It decreases monotonically down to 20\% at the highest energy. In contrast  the 3-fragment exit channel increases and saturates 
or even decreases at high energy. The 4-fragment exit channel appears at $E/A$ = 12 MeV and increases at highest energy, indicating the dominance of the multifragmentation decay mode. The crossing of the fission and 3-fragment exit channels is located between $E/A$ = 18 and 20 MeV, exactly at the maximum of the fusion cross section. One can speculate that the attractive nuclear potential overcomes the repulsive one at this energy. D. H. E. Gross\cite{gro93} has predicted this decomposition into different exit-channels for an excited $Au$ nucleus, using Microcanonical Metropolis Simulation of Statistical Multifragmentation (MMMC). Qualitative agreement with our result has been found. These experimental results provide crucial constraints for dynamical models as well.

\section{Estimation of primary source and secondary decay}
\begin{figure}[h]
\begin{minipage}{20pc}
\includegraphics[width=20pc]{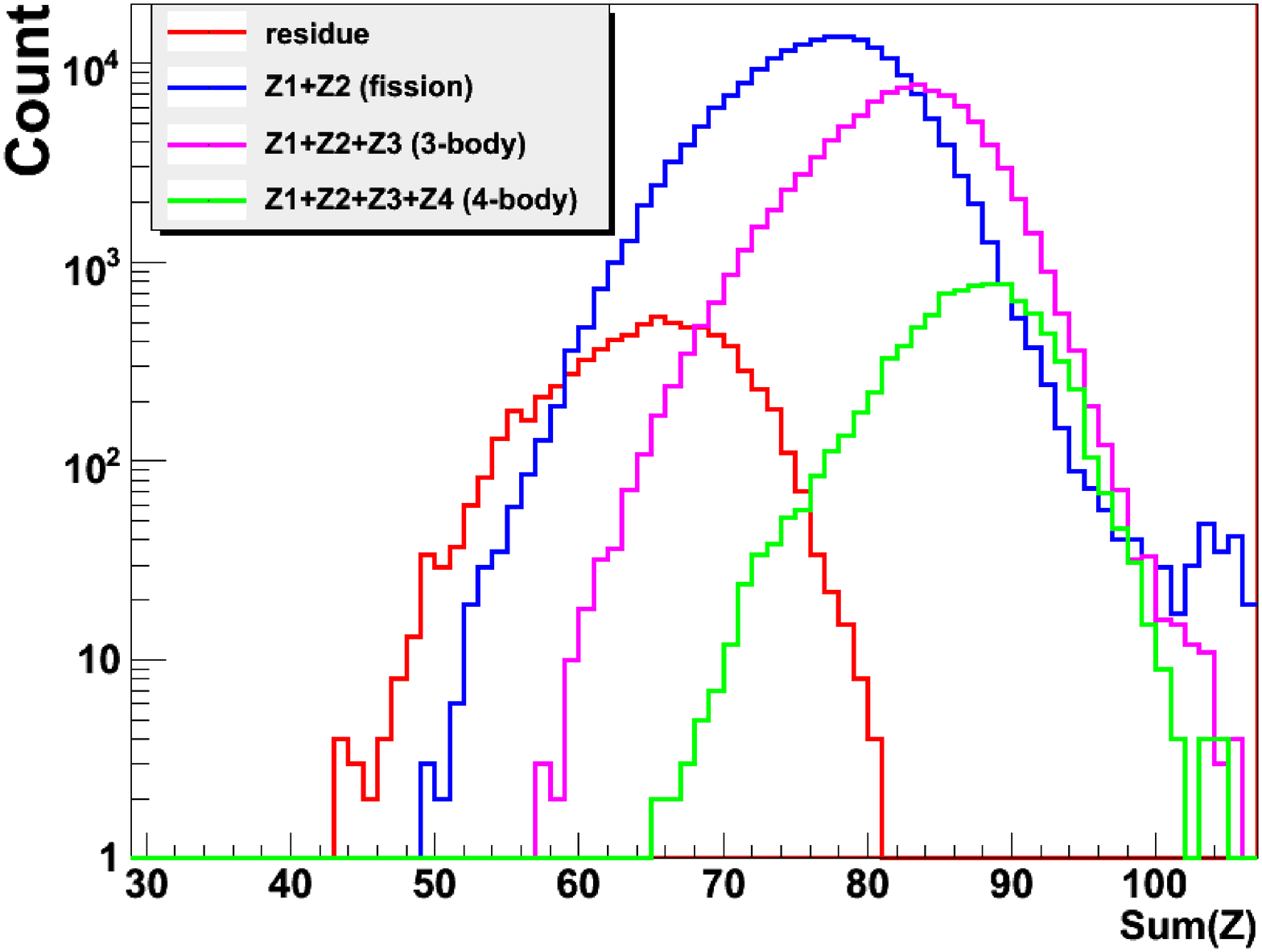}
\caption{\label{sumz}Charge sum, $Z_S$, of 2-, 3- or 4-fragment in the central collision of $^{129}Xe$+$^{nat}Sn$ at $E/A$ =  15 $MeV$.}
\end{minipage}\hspace{1pc}%
\begin{minipage}{20pc}
\includegraphics[width=20pc]{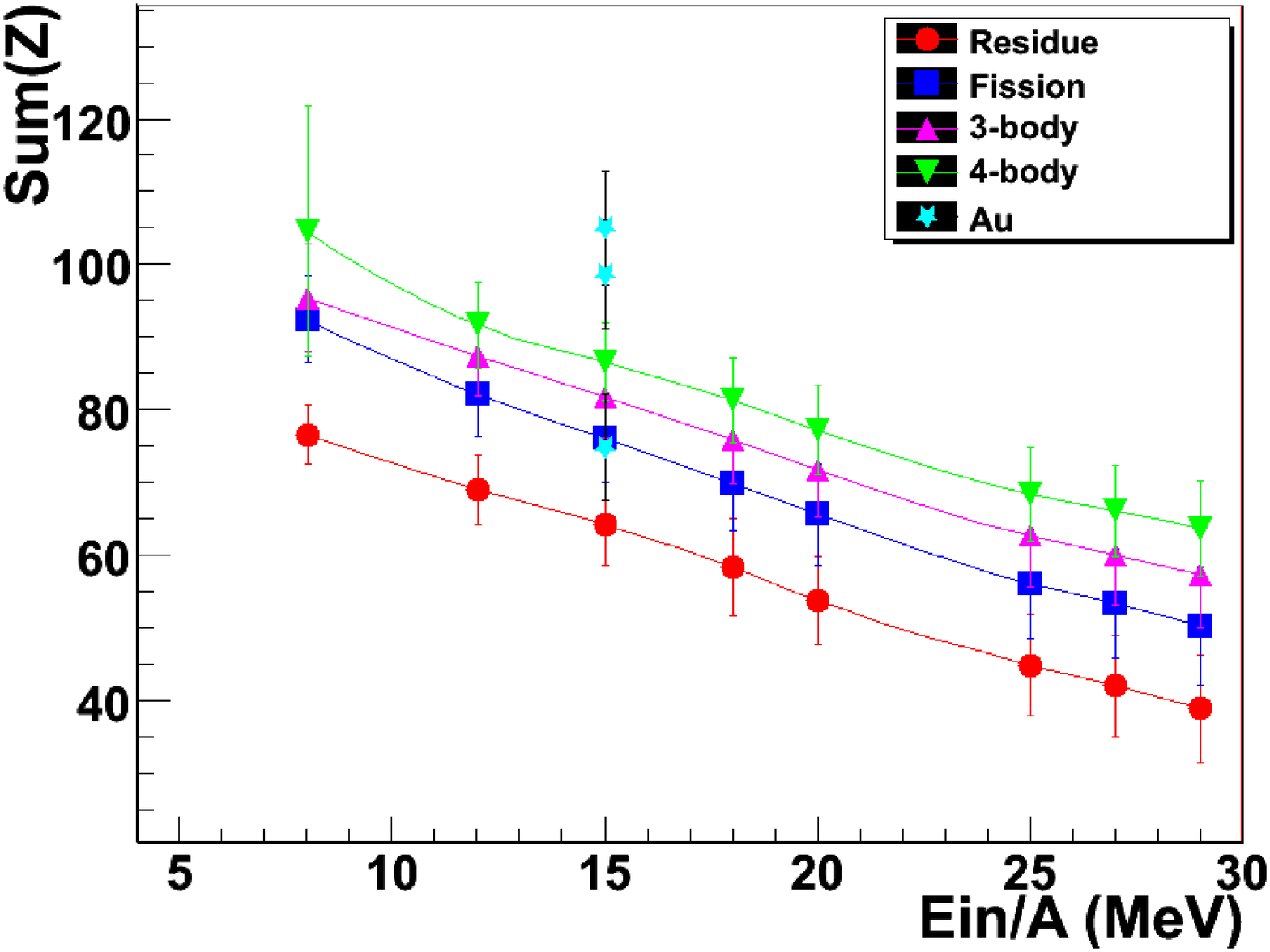}
\caption{\label{zs}Average value of $Z_S$ distribution in the central collisions of $^{129}Xe$+$^{nat}Sn$ as a function of the incident energy}
\end{minipage} 
\end{figure}

 Let us assume that the projectile and target have formed a composite system either by fusion or by massive transfer. Part of the projectile and target can be emitted as preequilibrium during this process. Then the highly excited composite system will decay by evaporation of light particles in competition with fission, 3-fragment or 4-fragment break-up. An estimation of the size of the composite system can be deduced from the sum of the charges of each exit-channel. The resulting charge of the source, $Z_S$, represents a lower limit. The distribution of $Z_S$  is presented in figure \ref{sumz} for  $^{129}Xe$+$^{nat}Sn$ at $E/A$ =  15 $MeV$. The charge distribution of surviving evaporation residues is also plotted in this figure for comparison.  The minimum size of the composite system can reach 90\% of the total system, in contrast with the size of the residue which reaches only 77\%. It increases with the number of fragments of each exit-channel. Figure \ref{zs} represents the average value of $Z_S$ distributions vs the incident energy. The average source size decreases monotonically with the increase of incident energy. 

Using the calorimetry method \cite{cus93} we have estimated roughly the excitation energy of the composite system. The contribution of all particles were considered in this estimation. The extracted excitation energies increase monotonically from $E^{*}/A$ =  1 $MeV$ at the lowest incident energy to $E^{*}/A$ =  5.5 $MeV$  at $E/A$ = 29 $MeV$.

\begin{figure}[h]
\begin{minipage}{22pc}
\includegraphics[width=22pc]{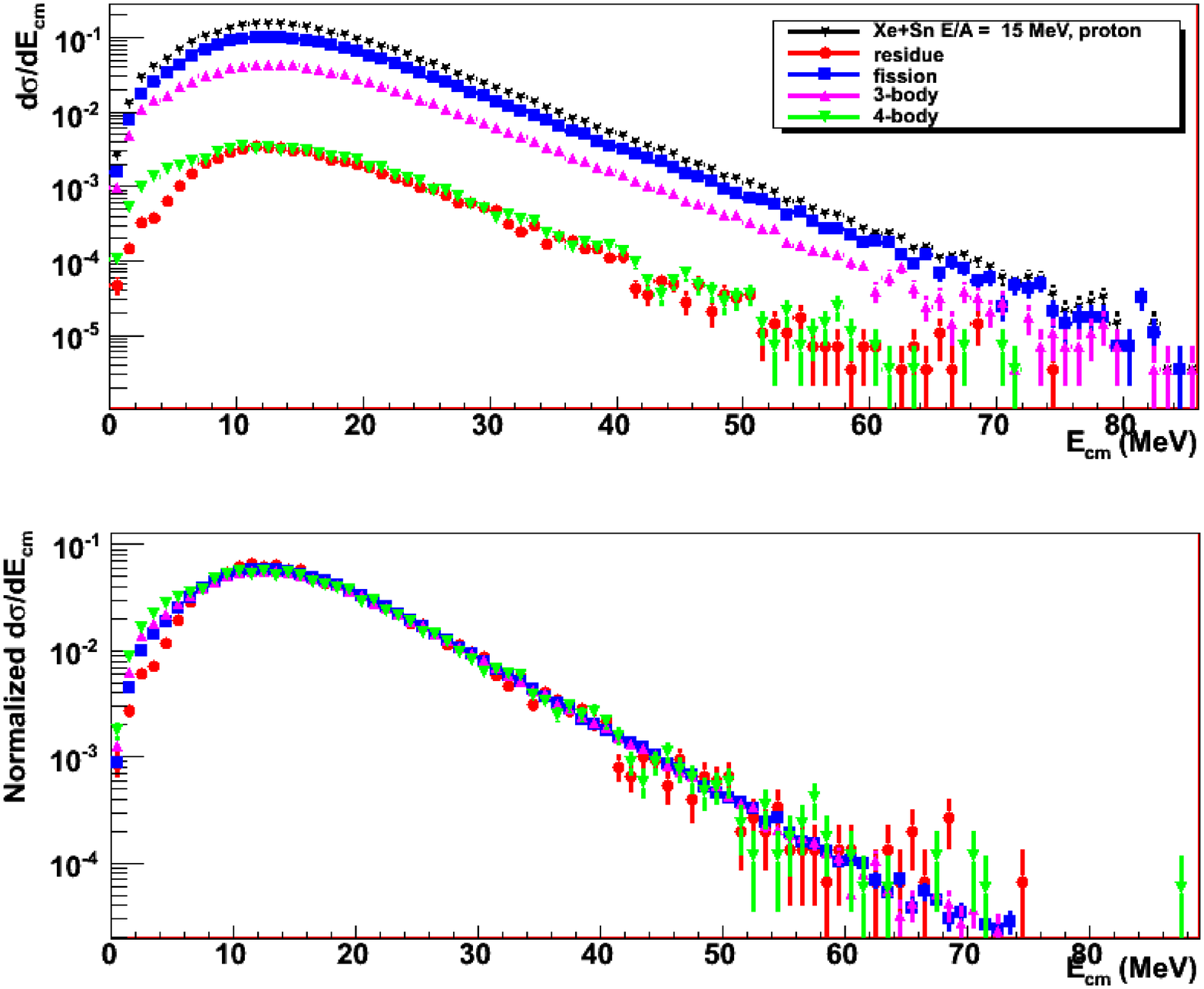}
\caption{\label{ep}Center of mass energy spectra of protons produced in coincidence with 1-, 2-, 3-, and 4-fragment for the central collisions of $^{129}Xe$+$^{nat}Sn$ at $E/A$ =  15 $MeV$. In the upper panel the spectra are relative to the total number of events while in the lower panel they are normalized }
\end{minipage}\hspace{0pc}%
\begin{minipage}{22pc}
\includegraphics[width=22pc]{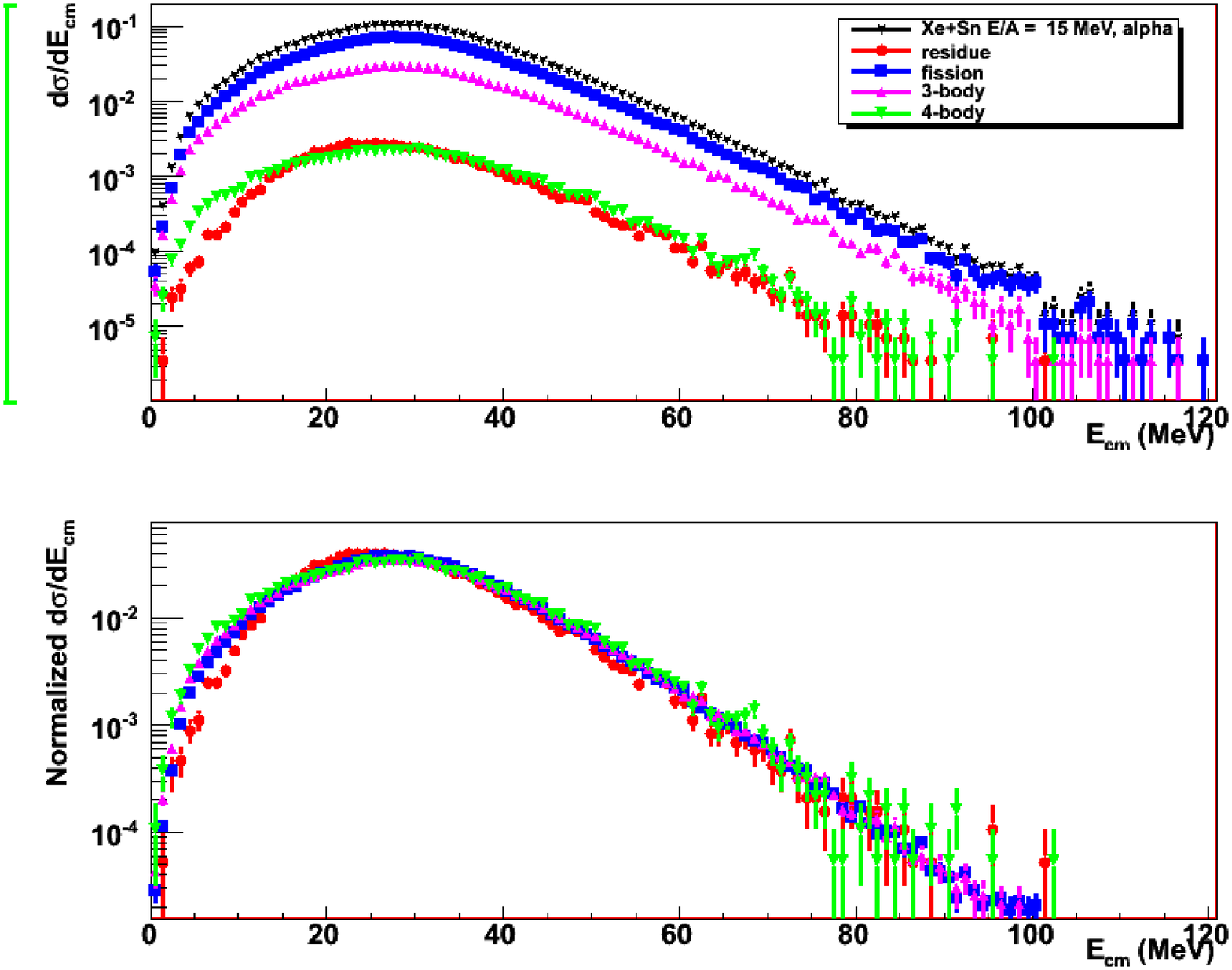}
\caption{\label{ea}Same as figure \ref{ep} but for $\alpha$ particles.}
\end{minipage} 
\end{figure}

Figure \ref{ep} and \ref{ea} represent the center of mass energy spectra of proton and $\alpha$ particles produced in coincidence with 1-, 2-, 3- and 4-fragment, respectively,  in  central collisions of  $^{129}Xe$+$^{nat}Sn$ at $E/A$ =  15 $MeV$.  In the upper panel of both figures, the measured spectra were divided by the total number of events (the integral of the distributions reflects their multiplicities) while in the lower panel they are normalized to unity. All spectra have the same Mawxellian shape with an overproduction at low energy ($E_{cm}\leq$ 10 MeV). This overproduction at low energy increases with the number of fragments. It can be interpreted as originating from secondary decay of fragments. The slope of the normalized spectra superimpose very well. This feature suggests that the particles are emitted from a common source having a defined temperature independent of the later break-up of that source in 2, 3 or 4 fragments .  
 
There are at least three different stages to produce the light particles:  in the early stage of the collision as preequilibrium particles;  particles evaporated from the compound system (primary emission);  they can be emitted from the excited 2-, 3 or 4-fragment (secondary emission). In case of the survival of the residue only the two first steps are considered.  

The energy spectra were fitted to a composite source model as it was done in refs.\cite{car03,keu04}. Three Maxwellian sources were included for fission exit-channel : one for the primary compound source and two for the fission fragments. The first fits did not provide a satisfactory representation of the spectra at all laboratory polar angles. The composite source model assumes an isotropic emission of the particles. However the excited compound system is highly rotating, causing an anisotropy in the emission of the particles. The azimuthal angles have to be taken into account to improve the method\cite{car03,keu04}.  These analyses are in progress and should give better results. 

An alternative method is to use correlation functions.  This is the method we employed to extract, on the average,
the LCP emitted from each fragment. With the help of intensive simulations, we have developed in refs.\cite{mar98,hud03} a correlation technique to extract the different contributions. In this earlier work it was possible to extract the secondary LCP emitted from the fragments produced in central collisions at intermediate energies. 

In the present work, we applied a similar but improved approach to extract the LCP emitted from the fission- , 3- and 4-fragment. The correlation function is defined as : 
\begin{equation}
1+R\left(V_{red}\right)=\frac{Y_{corr}(V_{red})}{Y_{uncorr}(V_{red})}.
\label{eqfctcorr}
\end{equation}
 where the correlated yield spectrum, $Y_{corr}(V_{red})$, is constructed with the fragment and LCP detected in the same event. This spectrum is sorted with respect to the reduced relative velocity, $V_{red}$, between the fragment and each LCP type. This quantity  is defined as  : 
\begin{equation}
V_{red} = 10. \frac{V_{rel}}{\sqrt{Z_{frag}+Z_{LCP}}}
\end{equation}
where $V_{rel}$ is the relative velocity of the fragment with charge $Z_{frag}$ and the LCP with the charge $Z_{LCP}$. This relation has the advantage to eliminate the charge dependence of the fragment-LCP relative velocity correlation functions ~\cite{kim92}. The uncorrelated yield spectrum, $Y_{uncorr}(V_{red})$, is constructed with an event-mixing technique ~\cite{zaj84,dri84}, consisting of taking the fragment from one event and LCP's from a different event. We also define the difference function $\Delta$$V_{red}$ as : 
\begin{equation}
\Delta V_{red} = Y_{corr}(V_{red}) - Y_{uncorr}(V_{red}).
\end{equation}
 
\begin{figure}[h]
\includegraphics[width=36pc]{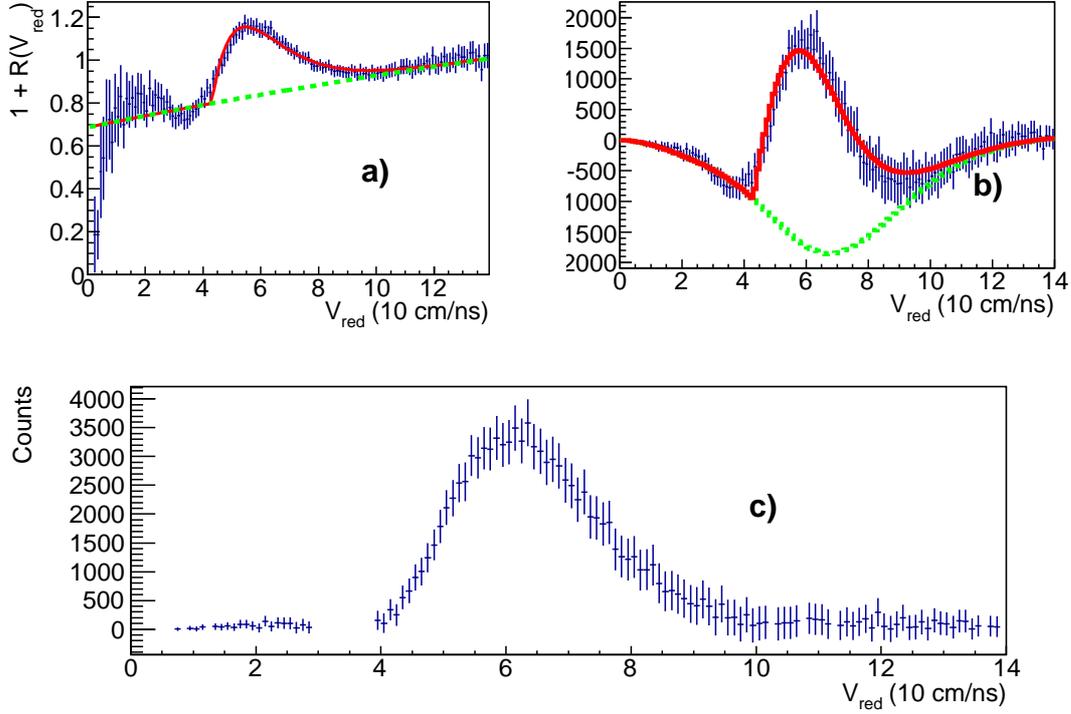}
\caption{\label{corfuncp}  Proton-fission fragment correlation function for the $^{129}Xe$+$^{nat}Sn$ at $E/A$ = 15 $MeV$. panel a) shows the ratio of correlated and uncorrelated events, panel b) the difference and panel c) represents the velocity of proton in the center of mass frame of the fissioning fragments, obtained from the subtraction of the difference function (data points) and the background (thick dashed curve).}
\end{figure}
\begin{figure}[h]
\includegraphics[width=36pc]{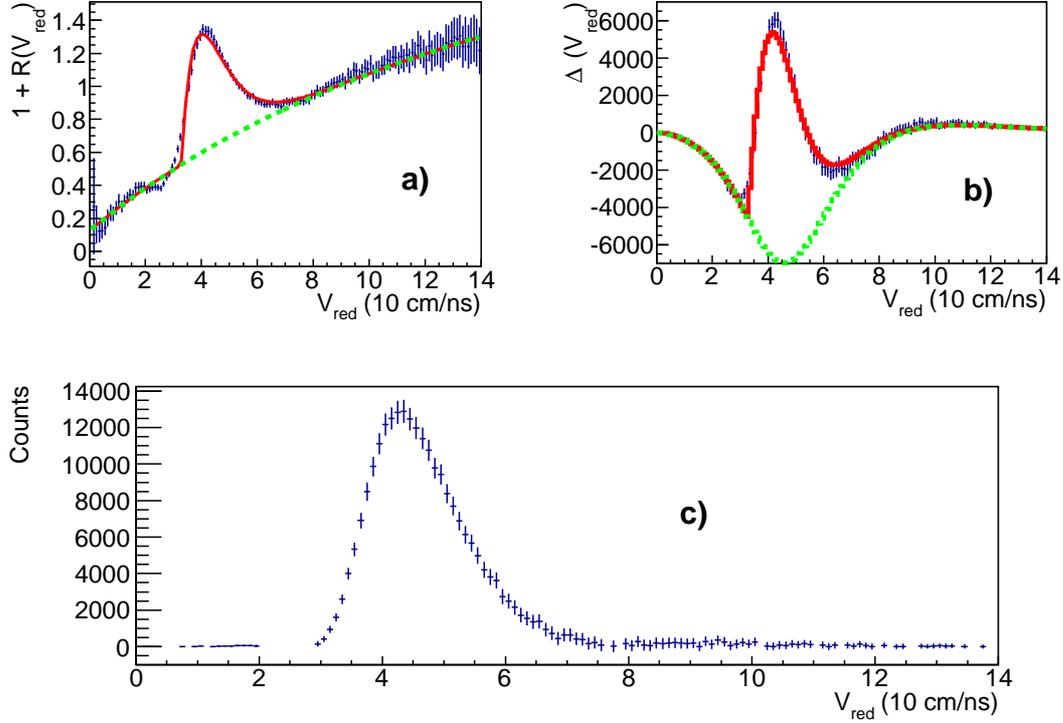}
\caption{\label{corfunca}  The same than figure \ref{corfuncp} but for $\alpha$-fission fragment correlations.}
\end{figure}

Figure \ref{corfuncp} shows a preliminary result of the correlation function (panel a) of reduced relative velocity of proton and fission fragments, the difference function distribution (panel b) and the extracted spectra of emitted protons  (panel c) for the  $^{129}Xe$+$^{nat}Sn$ at $E/A$ = 15 $MeV$ system. The same presentation is given in figure  \ref{corfunca} for the  $\alpha$ particles. The background has been parametrized by the functional : 
\begin{equation}
f_{bg}= A -  \frac{1}{B V_{red} + C}.
\end{equation}
where A, B and C  are free parameters. In order to determine the three parameters we fitted the correlation function with a sum of two functionals, one representing the background $f_{bg}$ and the second representing the signal to extract. The signal was parametrized by a  Maxwellian functional. The result of this fit is shown by thick curve in figure \ref{corfuncp}. We also superimposed the background curve ( thick-dashed curve). The velocity distribution of proton in the center of mass frame of the fissioning fragments is obtained by subtracting the difference function $\Delta V_{red}$  and the background function  $f_{bg}$ (transformed to the difference correlation function $\Delta$$V_{red}$). The result is shown in figure \ref{corfuncp} c) and  \ref{corfunca} c) . The multiplicity of secondary proton and $\alpha$  are deduced from the  integral of the distributions of panel c). 

 \begin{center}
\begin{table}[h]
\caption{\label{tab2}Total, secondary and primary multiplicities of light charged particles in coincidence with 1-, 2-, 3- and 4-fragment for the $^{129}Xe$+$^{nat}Sn$ system at $E/A$ = 15 MeV. The uncertainty of the secondary multiplicities is about 10\% }. 
\centering
\begin{tabular}{@{}*{7}{l}}
\br
Exit-channel & Particle & $M_{tot}$ & $M_{sec}$&$M_{pr}$ & $M_{sec}$/$M_{tot}$($\%$) \\
\mr
\verb"1-frag"&p &3.31& -  & -  \\
\verb"2-frag"&p &3.00&0.612&2.34&21\\
\verb"3-frag"&p &2.53&0.66&1.87&26 \\
\verb"4-frag"&p &2.14 &0.54&1.6&25 \\
\verb"  "& & &&& \\
\verb"1-frag"&d &1.06& - & - \\
\verb"2-frag"&d &0.95&0.33&0.62&35\\
\verb"3-frag"&d &0.79&0.32&0.47&41\\
\verb"4-frag"&d &0.65&0.36&0.29 &55 \\
\verb"  "& & &&& \\
\verb"1-frag"&t &0.60& - & - \\
\verb"2-frag"&t &0.51 &0.216&0.29&42\\
\verb"3-frag"&t &0.44&0.228&0.21&52\\
\verb"4-frag"&t &0.37& - & - \\
\verb"  "& & &&& \\
\verb"1-frag"&3He &0.15& - & - \\
\verb"2-frag"&3He &0.117 &0.0356&0.08&30\\
\verb"3-frag"&3He &0.093&-&-\\
\verb"4-frag"&3He &0.07& - & - \\
\verb"  "& & &&& \\
\verb"1-frag"&a &4.5& -  & -  \\
\verb"2-frag"&a &3.40&1.44&1.96&42\\
\verb"3-frag"&a &2.86&1.39&1.47&49\\
\verb"4-frag"&a &2.43 &1.56&0.87&64 \\
\br
\end{tabular}
\end{table}
\end{center}

The same procedure has been applied to all light charged particles for the four exit-channel when the statistic is sufficient. Moreover, we have used the multiplicities of secondary emission ($M_{sec}$) to deduce the multiplicities of the primary LCP emitted ($M_{pr}$) prior to scission or breakup of the compound system by subtracting them from the total multiplicities ($M_{tot}$).  

The total ($M_{tot}$), the extracted secondary ($M_{sec}$) and the primary ($M_{pr}$) multiplicities are given in table \ref{tab2} as well as the ratio of the secondary evaporation over the total emitted LCP's for the four exit-channel. First, the total and primary multiplicities decrease with the number of fragments. For protons, $M_{tot}$ decreases from 3.31 to 2.14 (20\%) and for $\alpha$ from 4.5 to 2.43 (30\%). This behavior indicates that part of the energy dissipation of compound system is removed by the breakup into 2-, 3- or 4-fragment.  

Second, no significant change in secondary multiplicities for a given LCP type is observed. The maximum change in $M_{sec}$ does not exceed 6\% which is bellow the uncertainty  of $M_{sec}$ (10\%). However, the production of secondary   $\alpha$'s is more than two time the production of protons. This observation can be explained by the high intrinsic angular momentum imparted by the heavy fragments which favors the emission of more complex particle such as  $\alpha$ particles. 

From the above observations, a possible scenario of the reaction mechanism is that the four exit- channel have a common history : 
i) A formation of highly excited and rotating compound system, which decays by emitted neutrons and LCP's (part of these particles can be emitted as preequilibrium). The value of primary multiplicities ($M_{pr}$) reflects most likely the time interval opened for their evaporation prior to fission or breakup into 3- or 4-fragment. The lower the $M_{pr}$ the shorter is the time interval to undergo breakup processes. The process to breakup into 4-fragment seems to start before the breakup into 3 and into 2 fragments, respectively. Some residues can survive the fission or breakup processes, but with low probability.
ii) After the breakup process, the formed fragments are excited and have a significant angular momentum. They decay by evaporation with a higher  $\alpha$ multiplicity than the proton one due to the high angular momentum involved. 

The analysis of the data for other beam energies is in progress. It should give a systematic of the secondary and primary multiplicities. Intensive simulations are currently underway and should allow time estimate of each step of the reaction.  
  
\section{Summary and Conclusion}
In this contribution we have presented the experimental results of the system $^{129}Xe$+$^{nat}Sn$ at $E/A$ = 8-29 MeV. The cross  section of the central collisions has been measured and reaches the value of 1.1 $b$ at  $E/A$ = 18-20 MeV. Very heavy residues are produced with significant cross section indicating the formation of heavy compound system either by fusion or by massive transfer processes. Four exit-channel have been identified consisting of 1-, 2-, 3- and 4-fragment. The relative yield of these exit-channels indicates that fission is the dominant decay mode at the lowest energy. The fission yield decreases progressively at the highest energy while the 3- and 4-fragment exit-channel increase. Similar probabilities are reached at $E/A$ = 18-20 MeV for 2-fragment and 3-fragment productions.  At this energy the attractive nuclear potential seems to overcome the repulsive  one. 

The size of the composite system has been estimated from the sum of the fragment charges. It can reach 90\% of the total system, in contrast with the size of the surviving residues which reaches only 77\%.

By using correlation functions we have estimated the total secondary emission of the LCP by the fragments for the  $^{129}Xe$+$^{nat}Sn$ at $E/A$ = 15 MeV. This secondary emission remains almost constant with the number of fragments in the exit-channel, while the primary emission (prior to the breakup of compound system) decreases.  This feature indicates that the process to breakup the compound system into 4-fragment starts before the breakup into 3 and into 2 fragments, respectively.      

\ack
We thank the staff of the GANIL Accelerator facility for their support during the experiment. This work was supported by Le Commissariat \`a l'Energie Atomique, Le Centre National de la Recherche Scientifique, Le Minist\`ere de l'Education Nationale, and le Conseil R\'egional de Basse Normandie.

\section*{References}
\noindent


\end{document}